\documentclass{aastex}
\usepackage{spr-astr-addons}

\RequirePackage{color}

\begin{document}
\newcommand{\amin}{$^{\prime}$}                   %arcus and coordinates
\newcommand{\asec}{$^{\prime \prime}$}
\newcommand{\adeg}{$^{\circ}$}
\newcommand{\adegdot}[2]{\mbox{#1$\stackrel {\circ}{_{\bf \cdot}}$#2}}
\newcommand{\amindot}[2]{\mbox{#1$\stackrel {\prime}{_{\bf \cdot}}$#2}}
\newcommand{\asecdot}[2]{\mbox{#1$\stackrel {\prime \prime}{_{\bf \cdot}}$#2}}
\title{Target star catalogue for Darwin\\ Nearby Stellar sample for a search for terrestrial planets}
\slugcomment{To appear in A\&SS}
%% Running heads
\shorttitle{Target stars for a search for terrestrial planets}
\shortauthors{Kaltenegger et al.}

\author{L.Kaltenegger} 
\affil{Harvard-Smithsonian Center for Astrophysics, MS-20 60 Garden Street, 02138 MA Cambridge, USA}
\and \author{C.Eiroa}
\affil{Universidad Autonoma de Madrid, Spain}
\and
\author{C. V. M. Fridlund}
\affil{ ESTEC/ESA, P.O. Box 299, NL-2200AG Noordwijk, The Netherlands}
%\email{\emaila}
%\altaffiltext{2}{Second Alternate Affilation.}
%\altaffiltext{3}{Third Alternate Affilation.}

\begin{abstract}
In order to evaluate and develop mission concepts for a search for Terrestrial Exoplanets, we have prepared a list of potential
target systems. In this paper we present and discuss the criteria for 
selecting potential target stars suitable for the search for Earth like planets, with a special emphasis on 
the aspects of the habitable zone for these stellar systems. Planets found within these zones would be 
potentially able to host complex life forms. We derive a final target star sample of potential target stars, 
the Darwin All Sky Star Catalogue (DASSC). The DASSC contains a sample of 2303 identified objects of which 284 are F, 
464 G, 883 K, 615 M type stars and 57 stars without B-V index. Of these objects 949 objects are flagged in the DASSC as 
multiple systems, resulting in 1229 single main sequence stars of which 107 are F, 235 are G, 536 are K, and 351 are M type.
We derive configuration dependent sub-catalogues from the DASSC for two technical designs, the initial baseline design and 
the advanced Emma design as well as a catalogue using an inner working angle cut off. 
We discuss the selection criteria, derived parameters and completeness of sample for different classes of stars.
\end{abstract}


%\section*{}
%\label{sec:intro}

\section{Introduction}

Since as far as we understand, life can only exist on planetary surfaces, in oceans, or in planetary atmospheres, the 
question if life exists elsewhere in the universe is closely coupled to the issue about the presence of planets orbiting 
around stars. This latter issue has been answered positively during the last decade. About 300 extrasolar planets 
orbiting stars other than the Sun have already been detected during the first productive 
13 years of searching\footnote{http://www.exoplanet.eu}, 
and hundreds, perhaps thousands more, are anticipated in the coming decades. The number of planets have been doubling about every 27 month since the early 90s.
Although planets are currently discovered at an accelerating rate, and increasingly at distances greater than 100 pc (e.g. through microlensing), most planets found are 
something more akin to the gas giant planets in our own solar system and significantly more massive than the Earth 
($>$ $5 M_{Earth}$ to $<$ $13 M_{Jupiter}$, the brown dwarf limit). This is in part due to selection effects, since the most successful methods for 
discovering individual planets are the radial velocity and the transit methods. From the ground both work best 
for larger planets. Many also orbit very close to their primaries and/or are found in very eccentric orbits -- something which is not thought to be benevolent to the emergence and evolution of life as we know it. 
The smallest extrasolar planets found so far orbit pulsars, and probably do not provide a good base 
to address questions about our own Solar System \citep{Wolszczan2004}. We expect planets where we will eventually find 
life to be small and rocky -- very similar to the Earth and although there exist other possibilities that have been 
theoretically explored, this idea focus the search very much towards systems like our own. The discovery of an 
extrasolar terrestrial planet is the goal of many different projects today, through which smaller and smaller bodies are being detected (e.g. Bennet et al., 	arXiv:0806.0025v1). Nevertheless, at the time of writing no 'true terrestrial analogues', i.e. an Earth size body in the 'Habitable Zone' has been reported.

Essentially all planned searches for such 'true terrestrial analogues', focus on this so-called 'Habitable Zone' (HZ, \cite{Kasting1993}), which is very loosely defined as the region around a particular star, where one would expect 
stable conditions for liquid water on a planetary surface (see section 3.4). The topic is very complex, since many factors 
unrelated to stellar luminosity could work to produce conditions where water would be liquid. As an example, the Jupiter moon 
Europa in our Solar system is expected to have liquid water under a layer of ice, where the energy to keep it liquid originates from 
gravitational interactions within the Jupiter system. A further complication would be to investigate all systems where one could imagine 
water to be liquid even very locally (scales less than 1 km), such as in the paper by \cite{Sagan76}, where the authors 
discuss possible ecologies in the atmosphere of Jupiter. The detection of such habitable conditions would be extremely difficult remotely. All planned searches for habitable planets thus focus on the HZ. This is also the approach of the Darwin and TPF studies -- as described below. 
A key issue in the search for and eventual study of terrestrial exoplanets is where, and around which stars they orbit. It is clear from 
the studies that have been performed so far, that any searches will have to eventually be carried out around nearby stars 
if we are hoping to be able to study any planetary bodies found in these searches. It is therefore crucial to analyze the 
potential targets in order to be able to estimate the chances of success. The refined target list also aids the construction of 
suitable instruments for the detection of habitable planets. 

This paper describes the first such process which has been carried out in the context of the 
European Space Agency's Darwin study. Section two describes the Darwin study and the space mission concept. Section three 
focusses on the selection criteria and the characteristics of the Darwin All Sky Target Star Catalogue (DASSC). Section four 
derives three sub-star-catalogues based on different technical designs and shows the influence of the these on the
target stars that can be sampled. Section five discusses target selection criteria in details and section 
six summarizes the conclusions of the paper. 

\section{The Darwin and Terrestrial Planet Finder Space missions}
\subsection{The Studies}
Proposed space projects such as Darwin \citep{Leger1996,Fridlund2000} and/or the Terrestrial Planet Finder (TPF, \cite{Beichman1998}) 
plan to search for and characterize terrestrial exoplanets. Here we will focus on the European Space Agency's (ESA) Darwin mission, 
which has been studied so far in a close collaboration with NASA's TPF-Interferometer (TPF-I). 
All conclusions in this paper apply also to the latter mission \citep{Beichman2007}.

The Darwin mission concept consist of a so-called 'nulling' or destructive interferometer launched into space. A nulling interferometer
combines the light from 
several individual telescopes, with appropriate phase delays, in such a way that the light along the optical axis of the 
synthesized beam is extinguished (\cite{Bracewell1978} \cite{BracewellMacPhie1979}). At the same time, the light from an object offset by a (very) small angle will be enhanced. 
The size of the angle depend on the distance between the telescopes \citep{kaltenegger2005}. By choosing this separation for 
a particular star, one can search the HZ for planets. The output of a nulling interferometer 
is an amplitude (usually time variable because of the modulation (\cite{kaltenegger2005})). By feeding this light through a 
spectrograph, the physical properties of the planet can be studied. For reasons of contrast, as well as suitable spectral 
lines, both the Darwin and the TPF-I studies selected the wavelength range between 
$\lambda$ = 6$\mu$m 
to 20$\mu$m for operations.

An instrument like Darwin will operate best for nearby stars. In principle an 
interferometer searching for Earth-like planets is {\it not} restricted by distance. However, although the baseline can be adjusted for the 
required resolution, the integration time will at some distance  be unrealistic for a given collecting area. An Earth size 
planet, located at a distance of 10 pc will have a flux of 0.34$\mu$Jy at 10 $\mu$m. Given that the collecting area of the studies will be undecided for some time,  we have chosen to limit our stellar target 
sample to a distance of 30 pc, representative of the largest instruments considered.

Since no a priori knowledge of the number of stars having terrestrial type planets exists, it is important that one will include as 
many stars as possible in any target catalogue. A detailed study of the target stars 
will take time, and since the catalogue is needed for the preparatory technology studies and detailed design work, the 
determination of an input catalogue has been initiated by ESA. 
To model realistic observation scenarios, the luminosity, distance and radius of each individual star and the corresponding HZ 
for Earth-like planets have to be taken into account. Detailed simulations for space based missions that search for earth-like 
exoplanets critically depend on correct data on the target stars. The resulting catalogue is presented in this paper together 
with selection criteria and constraints put on sub-catalogue by the technology of possible mission concepts.
\subsection{The status of the ESA study}
 The first theme in ESA's so-called Cosmic Vision 2015-2025 scientific plan is "What are the conditions for planet formation 
 and the emergence of life". As a consequence of this, in late 2007, ESA decided to continue studying the technology needed to enable the eventual study of Earth like exoplanets in Habitable Zones. At the time of writing, the planning of this process is ongoing. 
 
  As part of this enabling process, ESA has also decided to follow-up the successful French-European-Brazilian 
 CoRoT mission (See ESA SP1306, 2006), which is searching for transiting 'rocky' exoplanets at great distances, with a study of a 
 mission searching for transiting exoplanets around nearer stars. This mission is called 
 PLATO (PLAnetary Transits and Oscillations of stars) with a possible flight in 2017-18, and will provide valuable information that can be used for the actual 
 design of future missions like the Darwin study.
\subsection{Motivation}
Several lists of nearby target stars have been compiled  that are related to the topic of this paper, but intended for 
different purposes, e.g. for the SETI search \citep{Turnbull2003}, Terrestrial Planet Finder Coronagraph (TPF-C, \cite{Brown2006}), as well as the closest stars to the Sun e.g. \cite{Mello2006,Gray2003} respectively.
All of these groups have used criteria, different from what is required for an unbiased search such as by a Darwin type mission., 
Turnbull and Tarter (2003), for instance,  concentrate on evolutionary criteria of 
the targets sampled to establish a list of stars that if harboring planets, could potentially have evolved life that might transmit radio signals detectable by the SETI program. The TPF-C target star sample is optimized for reflected light and thus 
differs significantly from the presented sample for infrared detections, as explained later in this paper.

\section{Darwin All Sky Target Star Catalogue (DASSC)}

The Darwin target star catalogue presented in this paper is a coherent, large, distance limited sample that introduces as 
few biases (e.g. metallicity, age) as possible.

%__________________________________________________________________

\subsection{Description of the DASSC}

The preparation of a complete catalogue of candidates from which an operative subcatalogue would be selected - constrained by the actual technical implementation, has been carried out through an iterative process. Interim results from this effort 
have been previously reported \citep{Eiroa2003,Kaltenegger2004}. 

During the Darwin feasibility study \citep{Fridlund2000}, a preliminary list of target stars using the Gliese star catalogue 
as input was identified and used (Leg\`er and Ollivier, private communication). This compilation contained a small number of F, G and K stars. Based on the principles discerned from this 
early work we have now established 
an extended catalogue, but using instead the Hipparcos catalogue as our starting point to produce an unbiased target star sample.

\subsubsection{Parent Star sample}
For the very nearby stars we consider here, the Hipparcos data has typical parallax standard errors of about 1 milliarcssec (mas), 
which allows very precise distance measurements. It also includes accurate photometry (the  uncertainty in B-V is typically 
less than 0.02\,mag), and proper motion data, as well 
as information on variability and multiplicity \citep{hippcat}. 
We combine the Hipparcos information with data from the 2MASS catalogue \citep{2massCat} and the 
Catalogue of Components of Double and Multiple stars (CCDM \citep{CCDM}) and the ninth catalogue of spectroscopic binary orbits (SB9). 
Based on the combined data we derive characteristics like 
luminosity, radius, mass, effective temperature and the extent of the Habitable Zone for the preliminary targets for the star 
sample out to a distance of 30 pc from the Earth. The target star list presented was used for the trade off studies of 
different mission architectures for the proposed Darwin mission (see e.g. \citep{KalteneggerSPIE2004,Kaltenegger2006}  
for TPF-I configurations see e.g. \citep{Lay2005, Serge2004}).

\subsubsection{The main DASSC catalogue}
For the main DASSC catalogue, we selected all F G K and M stars within a distance of 30 pc (spectral classification of stars 
without an assigned spectral type is based on the B-V and V-K color index). We established their characteristics as well as 
derived several key parameters that are necessary for observations and instrument performance evaluation. This lead to the 
Darwin All Sky Star Catalogue (DASSC). It is a volume limited sample of Hipparcos stars that is magnitude-limited to about V-magnitude ($V_{mag}$)
of 7.5 \citep{hippcat} and includes stars to $V_{mag}$ 12 (the Hipparcos catalogue is however incomplete between $V_{mag}$ 7.5 and 12). 

A number of stars are flagged in the DASSC as being multiple. Systems that are given a multiple designation in the Hipparcos catalogue but not 
in the CCDM list are most probably spectroscopic binaries and introduce a bias in the data, since the derived parameters assume 
that the basic information is produced by a single star. Multiple systems found in the SB9 are spectroscopic binaries with orbital solutions.
All multiple objects are flagged in the DASSC in order to be able to either 
exclude them or treat them specially when producing configuration dependent sub-catalogues for models of flight hardware. 
We use a cutoff of $\pm$~1 mag from the Main Sequence to establish Main sequence character using Wright 2005 (see also Mamajek 2008) 
(see Figure 1). This removes 99 dwarfs and 185 giant stars from our list 
The resulting sample consists of 1962 identified objects of which 252 are F, 412 are G, 808 are K and 490 are M type stars. 
Of these objects 733 objects are given in the DASSC as multiple systems (see also below), leaving 1229 stars 
(of 1354 stars including giants and dwarfs) that are assumed to be single target stars. 
In total 949 of 2303 stars including giants and dwarfs are multiple systems.  
Figure 1 shows all single parent stars in the sample 
including giant and dwarf stars, given in table 3. The DASSC catalogue contains 1229 single main sequence stars of which 107 are F,
235 are G, 536 are K, and 351 are M type (incl 14 M stars with no B-V index, for these stars the calculations are based on the V-K index).

%%%%%%%%%%%%%%%%%%%%%%%%%%%%%%%%%%%%%%%%%%%%%%
% 
%                                                One column figure
%----------------------------------------------------------- Fig1
   \begin{figure}
   \centering
   \includegraphics[width=7.5cm]{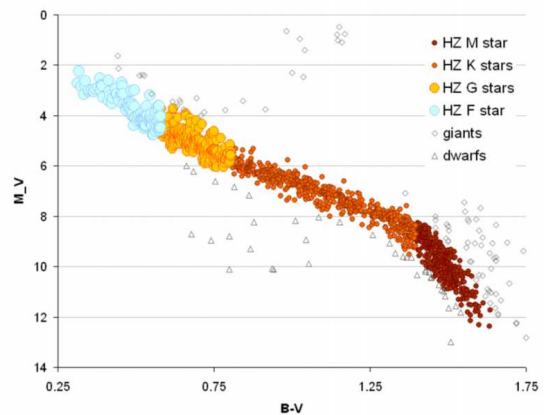}
      \caption{Color-magnitude diagram for the Single stars of the Darwin All Sky Target Stars Catalogue (DASSC).
              }
         \label{Fig1}
   \end{figure}

Table 1 (full table in electronic version only) shows the DASSC. It gives the HIP, HD, CCDM/SB9, distance, 
distance error, Spectral Type, B-V, V-K colors, V, J, H, K magnitudes, ecliptic coordinates, multiple flag and, in appropriate 
cases, Extrasolar Giant Planet (EGP) host star (status Nov. 2007). Table 2 (full table in electronic version only) gives derived quantities like luminosity, effective temperature, mass, radius, inner HZ radius, 
outer HZ radius and Johnson K magnitude. For completeness, Table 3 (full table in electronic version only) lists the remaining stars in the Hipparcos catalogue within 30pc 
(mainly White Dwarfs) with an outline like table 1. Table 4 (electronic full table by request only) lists the EGP host star and planetary system parameters of 
the systems within the DASSC (see http://www.exoplanet.eu). 
\subsection{Multiplicity}

Unsurprisingly, many of the preliminary targets are either spectroscopic binaries or 
have known resolved companions. Multiple stars are potentially interesting targets for searches for terrestrial exoplanets, since various groups have determined that stable orbits exist around multiple star 
systems. Further, \cite{Raghavan2006} found that about 30\% of Extrasolar Giant Planets (EGP) host stars are multiple. 
It still remains to be investigated exactly what are the constraints under which a given binary can or can 
not be observed with e.g. a nulling interferometer. A second star in the vicinity, especially within the field of view, induces a high background signal that inhibits the detection of an orbiting planet (this also 
applies to coronograph and occultation designs). For this reason, all known multiple systems have been removed in the first determination of 
instrument-depending sub-catalogues. We note that 716 out of the 2303 all sky targets are members of a double or multiple system tagged by CCDM. 
A further 211 additional multiple systems are flagged by Hipparcos, an additional 22 as spectroscopic binaries in SB9 \citep{sb9}.

 Multiplicity is a serious concern for the selection of target stars. A companion star in/near the HZ of a target star probably prohibits the formation of 
potentially habitable planets. The list of multiple systems in the DASSC is far from complete because not all multiple 
systems are flagged as being such in the CCDM catalogue (the CCDM catalogue only tags astrometric binaries with an apparent separation larger 
than about 0.1arcsec). Companions closer than 0.1arcsec are probably spectroscopic binaries (this will be addressed in a 
future paper). As a first step we have used the SB9 catalogue to exclude known spectroscopic binaries from the DASSC.

In the literature, we find the multiplicity fraction to be 57\% $\pm$ 10\% for F, G and K stars \citep{Halbwachs2003, Lada2006} 
and 26\% $\pm$ 6\% for M stars \citep{Delfosse2004}, respectively. In the DASSC we find 57\% multiple systems for F stars, 46\% for G stars, 35.5\% for K stars and
31\% for M stars (this excludes the stars with missing B-V index). The spectral classification is done by B-V color, a 
strategy adopted from \citep{Halbwachs2003}. Assuming a similar fraction of multiplicity for F, G and K stars, we estimate the number 
of missing binaries in the DASSC. G stars are missing 11\% binaries, K stars 21.5\%, which suggests that about 47 of 
the 247 G single DASSC stars and 190 of the 569 single DASSC K stars should be multiple systems. The F and M star multiple fraction 
is roughly as predicted within the error bars (see discussion of M stars Section 5.1.2).
The missing multiple systems are most likely spectroscopic binaries, most of them at separations smaller than 0.1 arcsec, which puts them potentially in or close to the HZ of the selected stars (see Figure 3). 

Some of the multiple systems flagged in the catalogue could be single stars because the CCDM catalogue also identifies apparent companions and not all stars designated as binaries because of stochastic solutions in Hipparcos are necessarily multiple systems,
since other effects could also lead to such solutions (G. Torres, private communication). Thus the list of multiple 
systems needs to be further evaluated. The DASSC single star list will be investigated to eliminate the existence of another 
star that can interfere with the nulling process (this number changes with the luminosity of the secondary star) or the 
stability of a terrestrial planet in the HZ of the target star. Also companion stars found outside the primary field of view 
of an interferometric design (the interferometric field of view is the same size as the point spread function of the individual telescopes) 
need to be modeled in order to evaluate the effect of their light on the background level. Not all multiple systems will 
degrade the performance of the instrument. The apparent separation at the time of the observation will be important to 
determine the suitability of the multiple system as a target. 

\subsection{Derived parameters}
We include the magnitude in J, H and K of stars in the DASSC for an easy assessment of planet to star contrast ratio. 
Assuming a solar metallicity, we  derive the effective temperatures of our 
star sample from the V-K index \citep{Alonso1996, Ramirez2005, Carpenter2001} for the conversion between the 2MASS, 
Johnson and CST systems. We assume solar metallicity for our sample since the results are relatively insensitive 
to the actual metallicity and only vary  by 30 K for $\pm$0.3\,dex \citep{Alonso1996}. Using the effective 
temperature we derive the BC \citep{Flower1996} and the applied corrections. We use the bolometric magnitude to derive 
the luminosity and radius of the stars in our sample shown in Table 2. We use the 2MASS catalogue (V-K) index to estimate 
the spectral class of the target stars with unknown spectral type, and to determine their Habitable Zone (see section 3.4). 

\subsection{The Habitable Zone}
 For a given planet (assuming a certain 
atmosphere composition and albedo) the surface temperature depends on the distance from the host star, the luminosity of 
the host star and the normalized solar flux factor $S_{eff}$ that takes the wavelength dependent intensity distribution 
of the spectrum of different spectral classes into account. The distance $d$ of the HZ can be calculated as \citep{Kasting1993}

\begin{equation}
d = 1 AU {\times ((L/L_{sun})/S_{eff})^{0.5}}
\end{equation}

where $S_{eff}$ is 1.90, 1.41, 1.05 and 1.05 for F, G, K and M stars respectively for the inner edge of the HZ (where runaway 
greenhouse occurs) and 0.46, 0.36, 0.27 and 0.27 for F, G, K and M stars respectively for the outer edge of the HZ (assuming 
a maximum greenhouse effect in the planet's atmosphere). These calculations were originally done for F0, G2 and M0 spectra 
and will be updated for all spectral sub classes (Kaltenegger, Segura and Kasting in prep). The size of the HZ  translates into 
instrument requirements like the minimum resolution required to detect the system. This is called the inner working angle in occultation and coronograph designs. Occultation and coronograph designs tend to have a set inner working angle based on their 
mask design while the maximum separation between the individual telescopes of an interferometer design sets the maximum 
resolution. A free flying interferometer design has the advantage that it is capable of resolving and thus detecting and 
characterizing planets very close to their stars - like habitable planets around the very large M stars sample in the solar 
neighborhood. Table 2 and Figure 2 show the radius of the host star and the HZ (equivalent to the Sun-Earth separation) 
in mas for the DASSC. Figure 3 shows the extent of the HZ and therefore what sub-sample we can probe for habitable 
planets depending on the 'inner working angle' or resolution of any instrument design. The extent of the HZ does not take the 
effect of clouds into account. The physics is, however, still very poorly understood, but the consequences should mainly be like: 
\begin{enumerate}
\item move the HZ toward the star if the cooling reflective properties of the clouds is stronger than their warming effect
\item move the HZ outwards if the warming effect is stronger than the cooling, or
\item potentially extending the HZ in both directions
\end{enumerate}
if the water clouds that form at the inner edge of the HZ have a cooling while the $CO_2$ clouds that form at the outer edge of the HZ 
have a warming effect. 
%
%                                                One column figure
%----------------------------------------------------------- Fig2
   \begin{figure}[t]
   \centering
   \includegraphics[width=7.5cm]{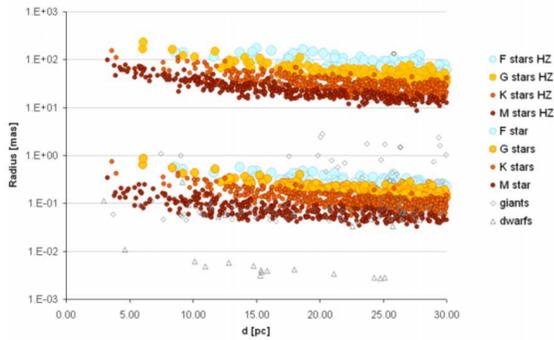}
   \caption{Radius of the star and the habitable zone in mas for the DASSC. 
	      }
         \label{Fig2}
   \end{figure}

%                                                One column figure
%----------------------------------------------------------- Fig3
   \begin{figure}
   \centering
   \includegraphics[width=7cm]{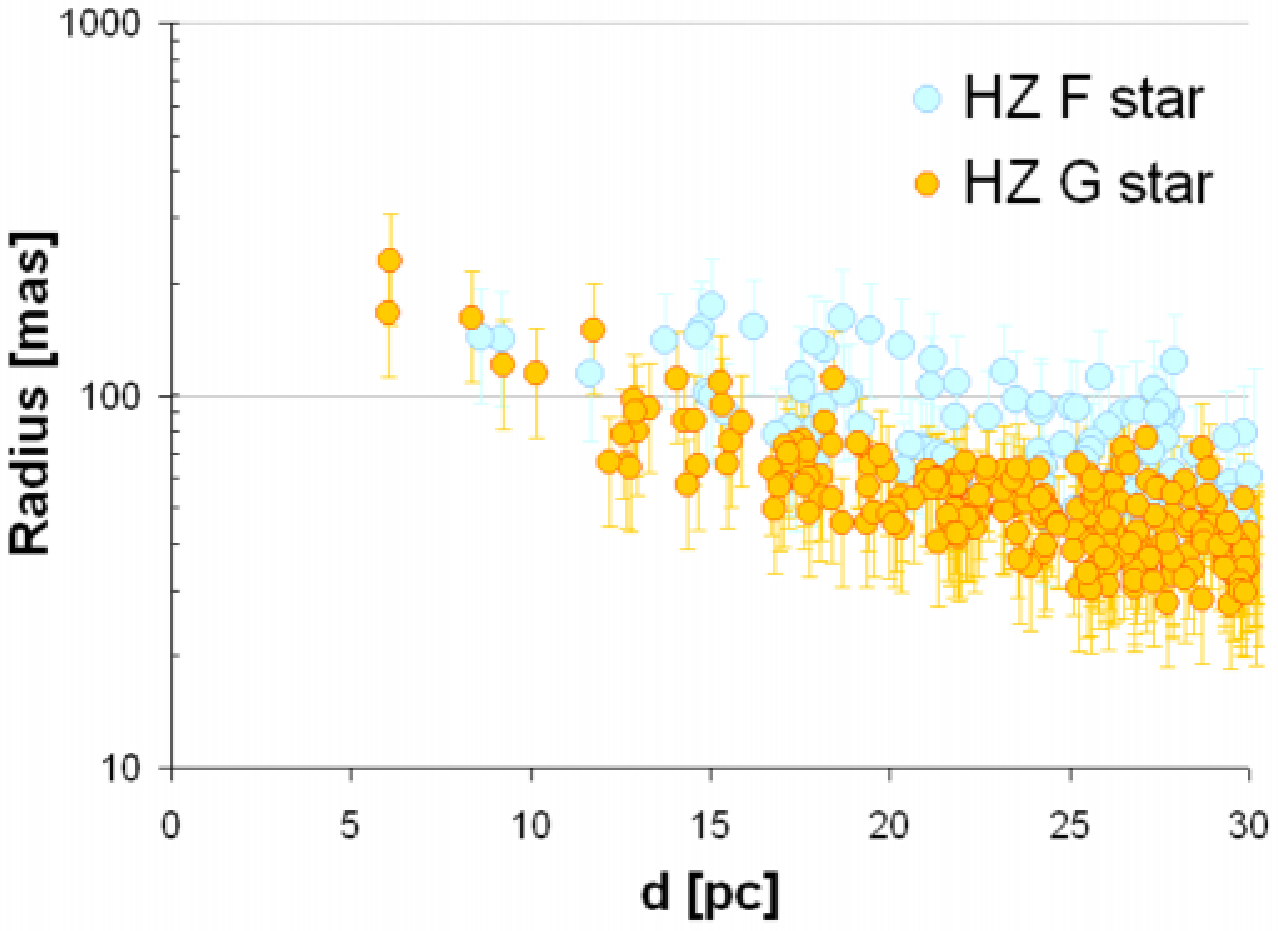}
   \includegraphics[width=7cm]{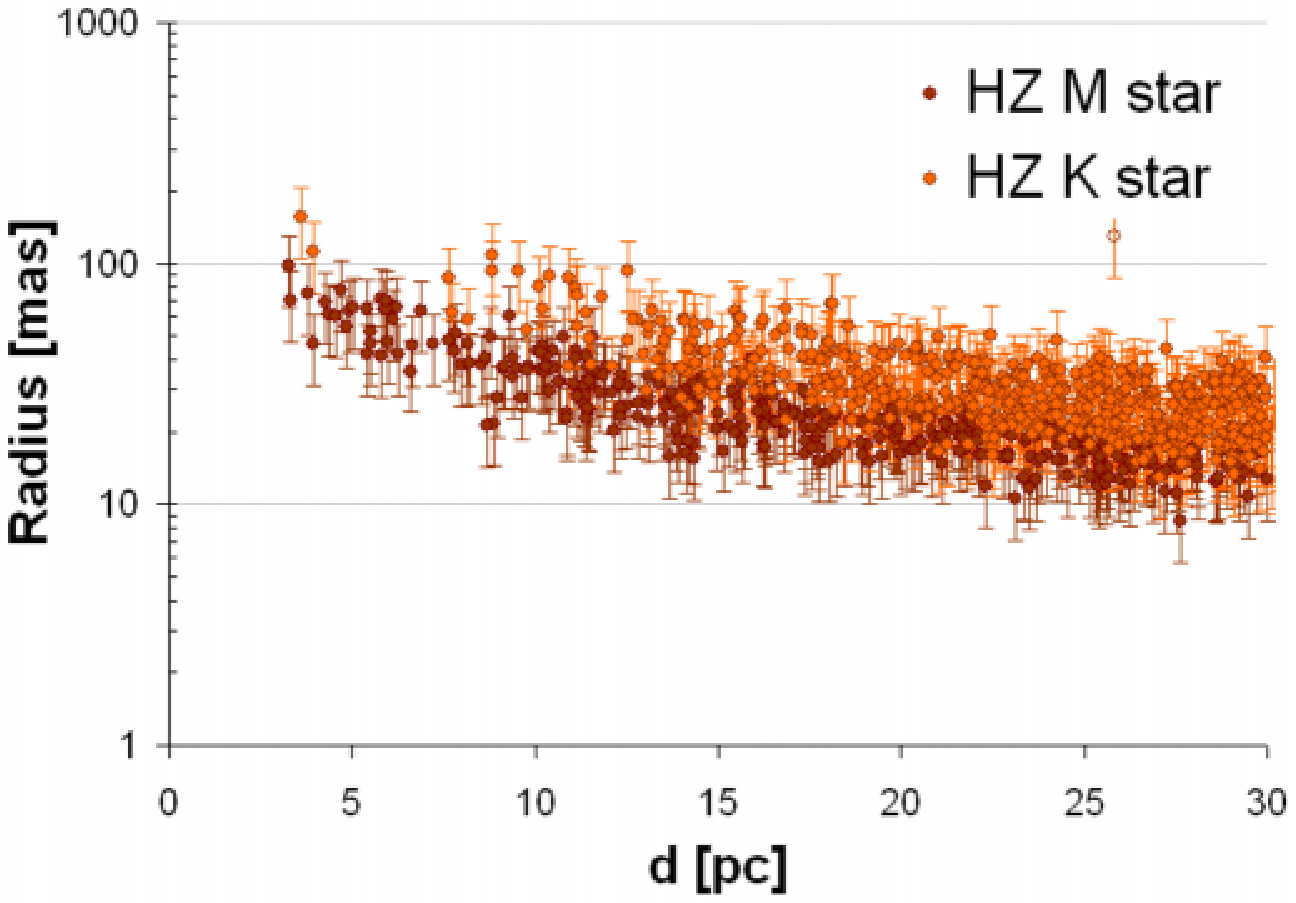}
      \caption{Extent of the habitable zone in mas for a) F and G stars and b) K and M stars in the DASSC.
            }
         \label{Fig3}
   \end{figure}

\subsection{Completeness of the All Sky Catalogue}
The Hipparcos database is approximately magnitude limited, but for later stellar classes, as well as densely populated areas 
it is incomplete \citep{Scalo2007}. We used the RECONS star list \footnote{www.chara.gsu.edu/RECONS/} of stars within 10pc to evaluate the completeness of our 
target star 
sample by comparing the number of stars within 10 pc in the DASSC with the Recons star list. The comparison shows that the F and G star 
sample is complete, the K star sample is at 73\% while the M star sample is only 45\% complete. 

\begin{figure}[t]
   \centering
    \includegraphics[width=7.5cm]{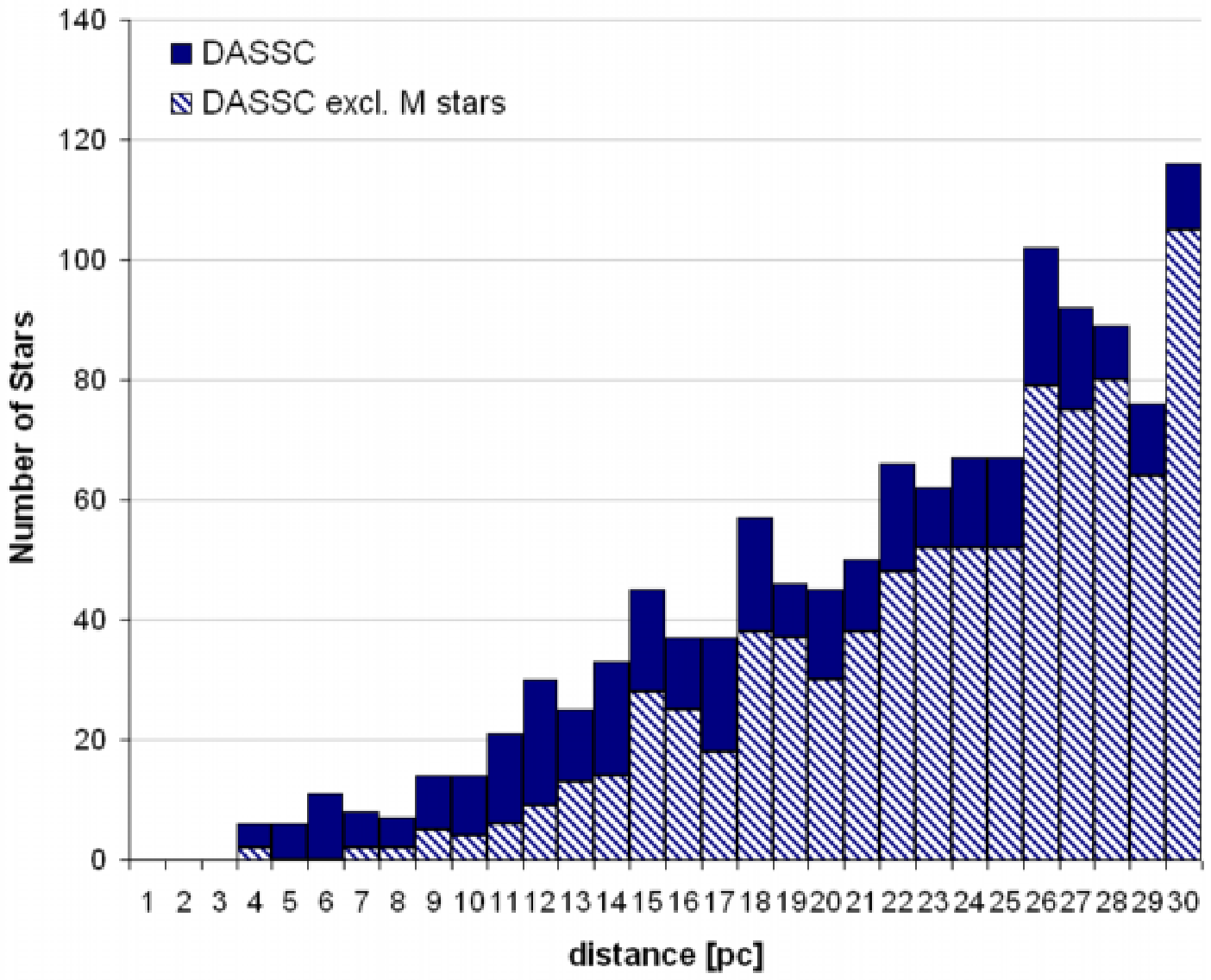}
       \caption{Number of all sky Darwin target stars versus 
      distance (solid color: all stars, stripes: excluding M stars).
     }
         \label{Fig4}
   \end{figure}

Using the number of DASSC stars within 10 pc we estimate the completeness versus distance of the DASSC: at 20 pc the catalogue is 
80 \% complete while at 30 pc it is only 53 \% complete. This number is heavily influenced by the missing M stars. 
Excluding the M stars from the estimate leads to a high completeness at 20 pc and 30 pc of 97 \% and 76 \% respectively, 
based on the estimated number of stars per volume using the 10 pc RECONS data set. The slight overestimation at 20 pc can 
be attributed to a) small number statistics and b) classification of K versus M stars that is based on a B-V color index 
being less or equal to 1.4 for a K- and larger for a M star.  
Fig.~\ref{Fig4} shows the number of DASSC stars versus distance.

To provide a homogeneous, sample of target stars, we choose not to include data on M stars from individual surveys 
(see e.g. \cite{Reid2004}) in the DASSC. Such surveys could provide additional low luminosity M dwarfs to the sample. 
This biases the DASSC towards the blue end. ESA is developing the GAIA astrometric survey mission for a planned launch in 2012. 
This mission is expected to provide parallaxes down to 200 $\mu$arcsec for objects brighter than R-magnitude 22 within 1.5 years 
of mission time (eventually during this mission, a precision of 20 $\mu$arcsec will be reached). Thus, we expect that this will 
complete our knowledge of nearby stars out to at least 30 pc and for essentially all stars of class K and M. The eventual goal 
is thus to have a complete homogeneous catalogue for all main sequence stars down to and including class M, out to this distance.
We note in this context that one of the purposes in preparing the current catalogue is also to provide input for supporting preparatory 
observations that will induce a further in depth characterization of the target star sample. For the late K 
and M stars, data on multiplicity, precise stellar type, rotation and metallicity are very incomplete. Thus 
modifications of the Darwin catalogue will be implemented as such data becomes available in a continuous process.
%
%                                                One column figure
%----------------------------------------------------------- Fig4
The important point concerning the limiting distance of the host star is the flux level we expect to receive from the planet. An Earth-like 
planet in the HZ is assumed to have a T$_{eff}$ of 300$\pm$ 50 K, which will determine its IR flux. In the visible the planets reflects starlight, a fraction of which, depends on the stellar luminosity, the star-planet 
distance and the planet's albedo. The 30 pc distance limit is imposed by the planetary flux that will be observable with reasonably sized instruments, and is not a limit due to the apparent star-planet separation: A planet in the HZ of an M star has a smaller apparent 
angular separation at 30 pc, than a planet in the HZ of an F star at 30 pc. IR free flyer interferometers can adjust 
the baseline of the configuration to the individual target star observed. 
\section{Sub-catalogues derived for study specific architectures}
Additional selection criteria will very likely be added because of constraints caused by the design and actual flight 
configuration of the instrument. Depending on the specific design, different parts of the sky will be accessible and 
sub-catalogues need to be derived for these. The actual target star list of a given sub-catalogue can also  
influence the choice of technical implementation. As an illustrative example, the original designs 
for the Darwin study assumed that the separate telescopes making up the interferometer are deployed in such a way 
that scattered (Sun-)light will not enter the light paths. When the individual telescopes and the beam combiner are 
all situated in the same plane, this implies that the optical axis of the synthesized beam of the interferometer can 
only be turned a maximum of $\pm$ 45\adeg from the anti-solar point along the ecliptic. A configuration developed 
later (the so-called 'EMMA' architecture) where the beam combiner is flying at a large distance away from the plane 
of the light collecting telescopes, instead, allow a pointing circle of $\pm$ $\adegdot{71}{8}$ around the anti-solar point. 
While this appear to offer a greater sample of target stars, 
nevertheless the issue of greater technical complexity and other observational technicalities need to be traded off 
before a choice of either of these (or perhaps even more efficient architectures) is selected. We have flagged the 
accessible stars of the two mentioned sub-catalogues ($\pm$ 45\adeg, and the EMMA) in the DASSC, and below we describe 
the consequence of selecting either of these sub-catalogues as well as one sub-catalogue that is based on an inner 
working angle limit of 75 mas. 
\subsection{The Darwin $\pm$ 45\adeg sub-catalogue}
This configuration is a design that allow to access $\pm$ 45\adeg along the ecliptic plane. 
The $\pm$ 45\adeg constraint discussed 
in this section derive from the actual spacecraft mechanics/design, of the initial version of Darwin \citep{Fridlund2000}. 
It is only one of several and depend on the actual configuration  chosen ultimately. It is discussed here for mainly 
two reasons: 1) It is a minimum selection that provides enough objects to fulfill mission requirements 
2) It provides a nice model to demonstrate how these criteria affect scientific objectives (which is our main concern in 
this paper) and can thus guide further refinement. We find that 
   \begin{itemize}
        \item 888 of the 1229 single all sky target stars are inside of the $\pm$ 45\adeg cone, forming this particular target list, 
	\item 281 of the 888 target stars are M stars, 379 K stars, 158 G stars and 70 F stars
        \item 25 of the 888 stars are known to host exo-planets.
   \end{itemize}
\subsection{The EMMA $\pm$ 71.8\adeg sub-catalogue}
 The EMMA configuration is a design that allows to access $\pm$ 71.8\adeg star sample. We find that 
   \begin{itemize}
        \item 1178 of the 1229 single all sky target stars are inside of the $\pm$ 71.8\adeg cone, forming this particular target list, 
	\item 343 of the 1178 target stars are M stars, 514 K stars, 217 G stars and 103 F stars 
        \item 36 of the 1178 stars are known to host exo-planets.
   \end{itemize}

\subsection{Inner working Angle 75 mas sub-catalogue}
This selection shows how an inner working angle influences the star sample. 75 mas was picked as a realistic value for 
some coronagraph designs. We used the equivalent Sun-Earth separation as the cutoff for the stars in this sub-catalogue. 
If the whole HZ (including the inner edge of the HZ) should be probed, that would require an inner working angle smaller 
than 75 mas for the stars selected - to find a planet.We find that 
   \begin{itemize}
        \item 89 of the 1229 single all sky target stars have a radius of the HZ bigger than 75mas
	\item 3 of the 89 target stars are M stars, 12 K stars, 21 G stars and 53 F stars.
        \item 10 of the 89 stars are known to host exo-planets
   \end{itemize}
 One should note that the sample is significantly smaller for the occulters/coronographs. This is of course due to that the number of stars where the HZ is inaccessible rises rapidly both with distance and with the intrinsic faintness of the later type stars.

%                                                One column figure
%----------------------------------------------------------- Fig5
   \begin{figure}[t]
   \centering
   \includegraphics[width=7.5cm]{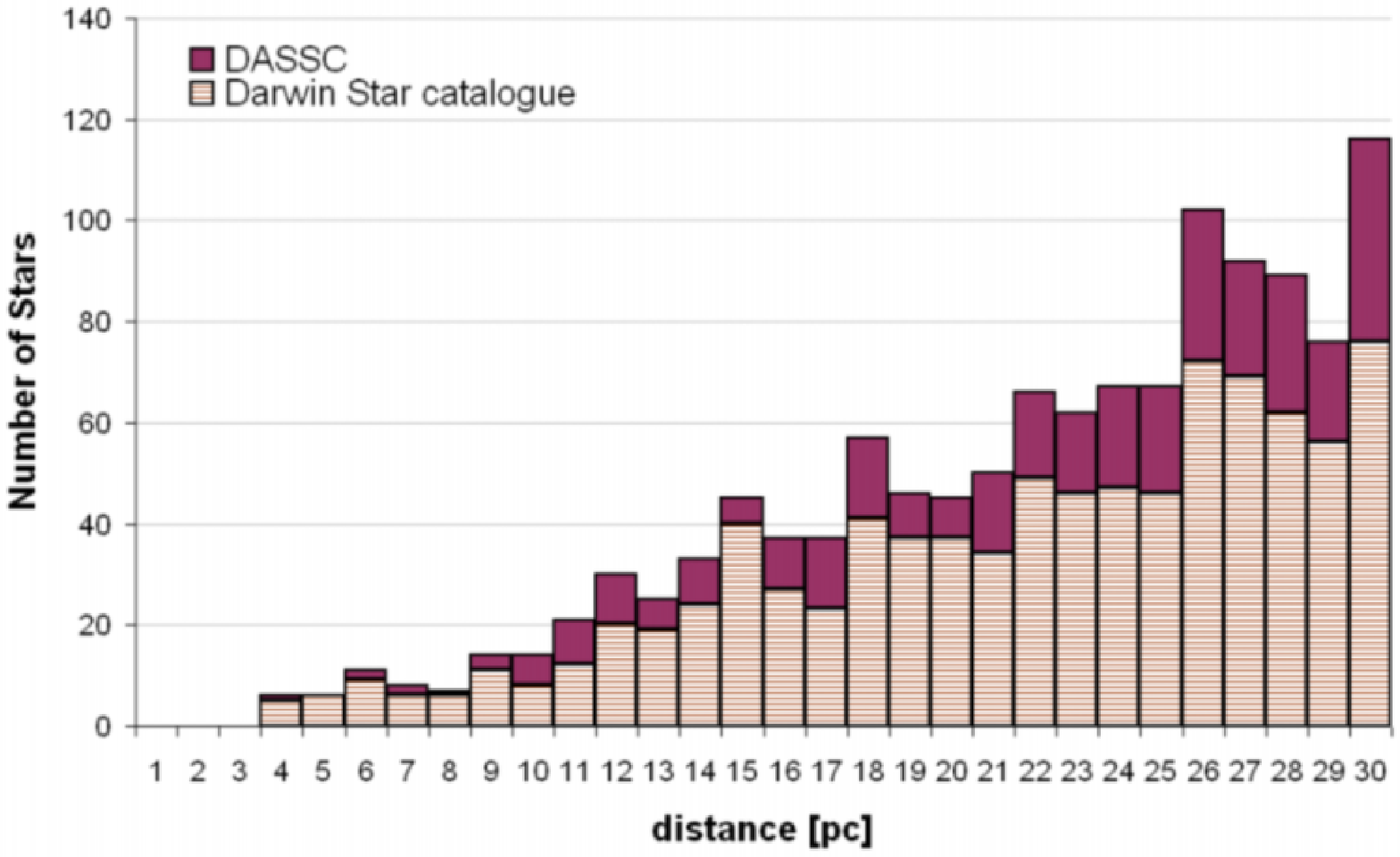}
   \includegraphics[width=7.5cm]{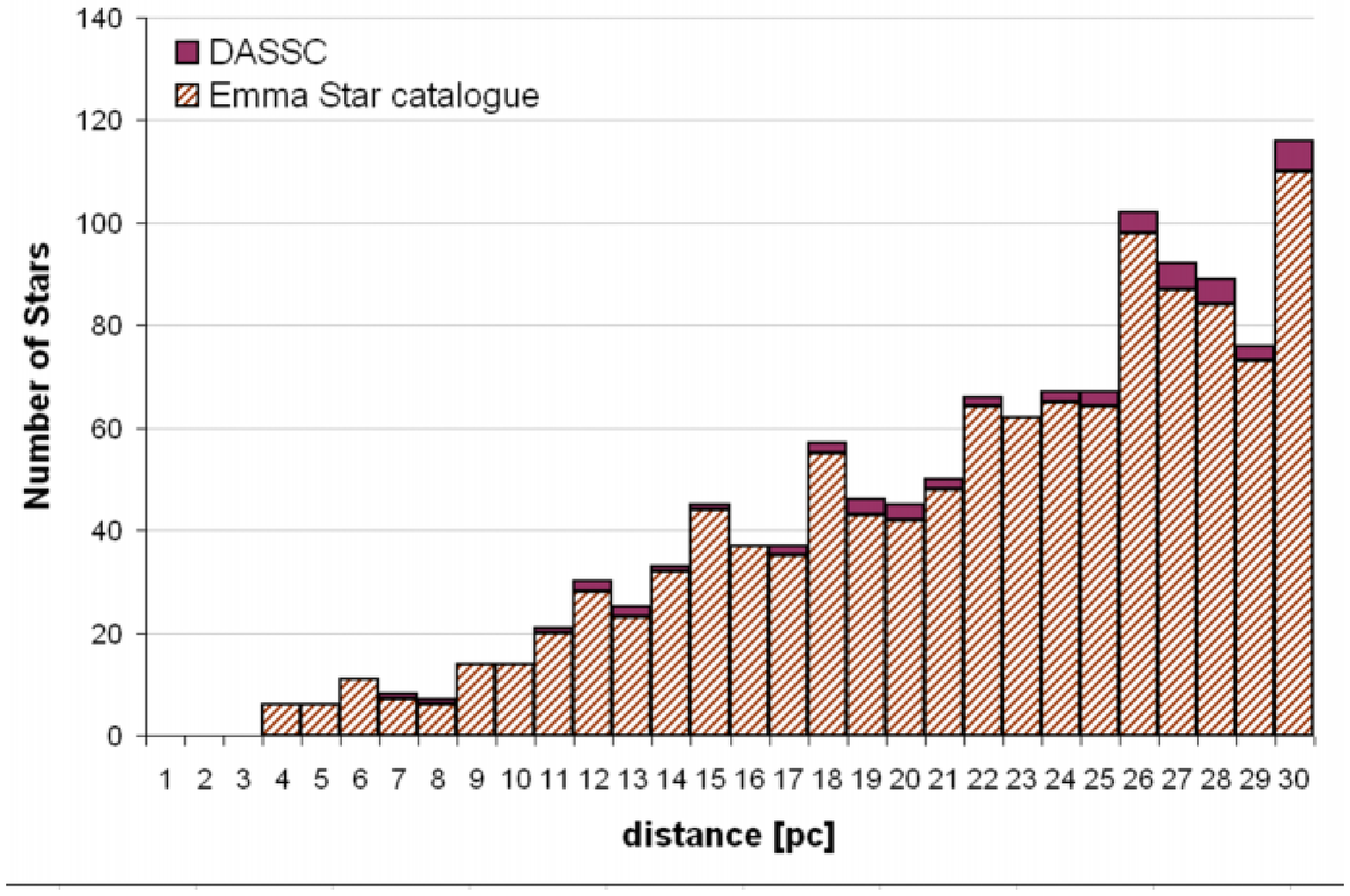}
   \includegraphics[width=7.5cm]{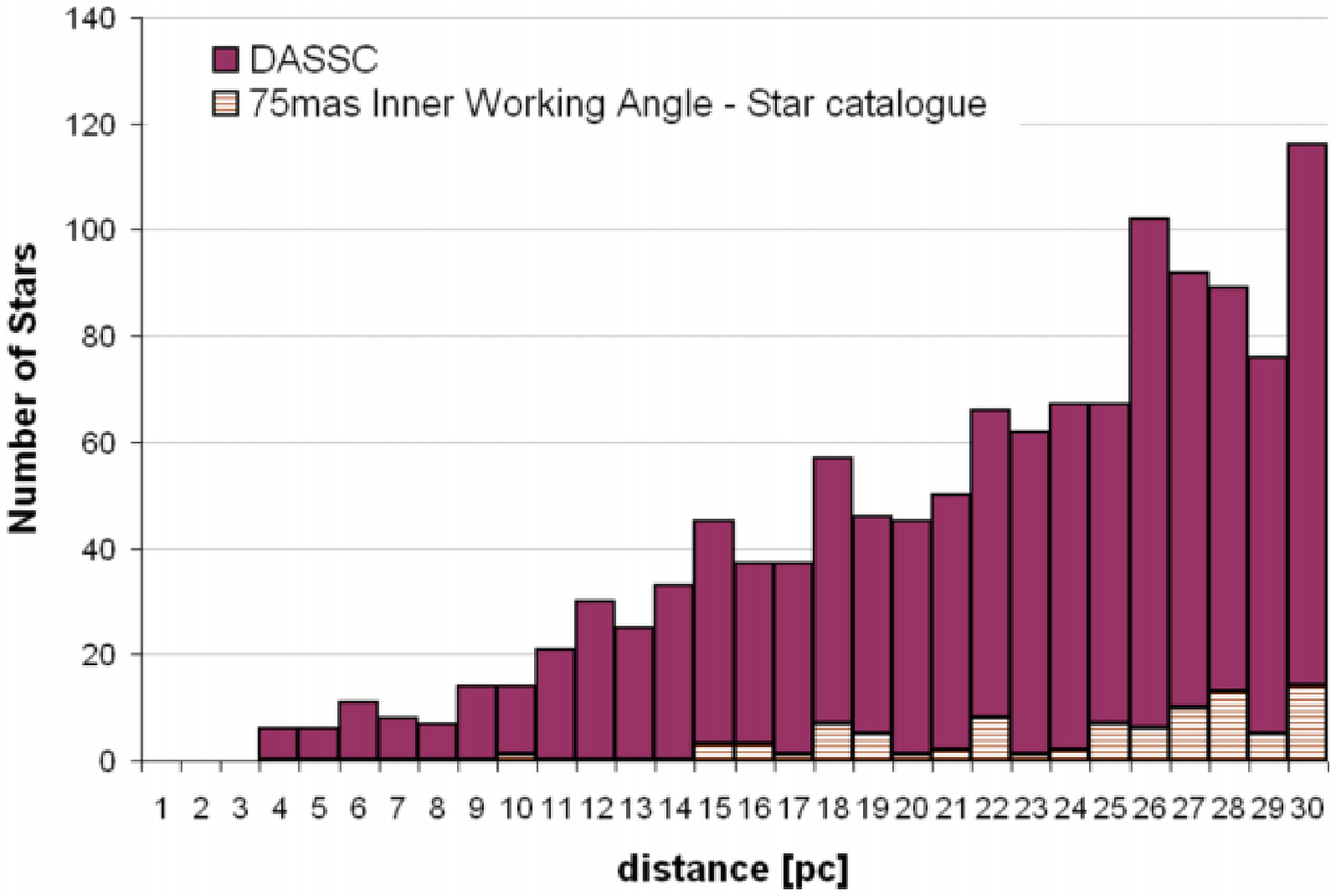}
      \caption{Number of DASSC target stars versus distance for the three discussed configurations a) the initial $\pm$ 45 degree cone 
      and b) the Emma $\pm$ 71.8 degree cone configuration c) 75mas cutoff inner working angle (solid color: all sky targets, stripes: sub-catalogues)
              }
         \label{Fig5}
   \end{figure}

\section{Discussion}
\subsection{Issues for target star selection}
\subsubsection{Spectral considerations}
 G stars are historically considered to be the prime targets for a search for habitable planets. However, the results of 
 exo-planetology and astrobiology during the last decade implies that a wider selection of targets stars should be made. 
 Since planets have now been found around essentially all different stellar types between class F and M, the target sample 
 should reflect this. Further, considering habitability, it is now generally thought that the main issue is the ability of 
 water to remain liquid on the planet's surface. This requires a relatively dense atmosphere. There is nothing a priori 
 excluding any spectral type from the catalog based on this requirement. This includes all types having a main sequence 
 life time long enough to allow the processes known to have occurred on the Earth to take place, {\it viz.} F - M. 
For a detailed discussion of habitability of earth-like planets around host stars of different spectral type (F,G,K and M) 
and detectable biomarkers in their atmosphere for the Darwin mission, we refer the reader to \cite{Segura2003}, \cite{Segura2005}, \cite{Selsis2002}, and 
\cite{Scalo2007}. The evolution of biomarkers in the Earth's atmosphere during different epochs have been investigated 
by \cite{Kaltenegger2006a} in respect to instrument requirements. The important issue is of course that research into 
evolutionary sequences of terrestrial planets will require sequences of stellar ages in the catalogue sample 
(Eiroa et al in prep).

  \subsubsection{M stars}
Especially interesting in this context are the red dwarf stars of class M. They are the most common of stellar objects and are thus found in great abundance close to the Earth. Nevertheless, they have been questioned as possible host stars for habitable planets because of their faint luminosity, requiring a planet to be very close to the star in order to be within the HZ. This latter aspect may force the planet to have bound rotation 
  \citep{Dole1964}, which could lead to processes where key gases freeze out on the dark side. Red dwarf stars also are prone 
  to flare activity orders of magnitude higher than anything our Sun has experienced since the T-Tauri stage prior to the epochs when life arose on the Earth. Such activity could inhibit the emergence of life and its 
  subsequent evolution. Further, the high levels of particle flow associated with strong flares could strip significant 
  amounts of atmosphere from a terrestrial planet irrespective of the strength of the planets magnetic field (Lammer, private 
  communication). On the other hand, it has been demonstrated, through detailed modeling, 
\citep{Joshi1997,Joshi2003} that even rather thin atmospheres could be sustained without freezing out. Many issues remain to be clarified, but at least planets have been discovered orbiting red dwarfs 
\citep{Udry2007}. To observe Earth-like planets in the HZ around a given star, the thermal flux will to first order be constant for a given planetary size, while the 
reflected stellar flux will scale with the brightness of the star. The suppression of the primary's thermal emission will, on 
the other hand, be progressively easier for later and later spectral types. For main-sequence stars, the star to planet contrast ratio is a factor 
of about 3 more for F stars, 3 less for K stars and about 11 less for M stars compared to the sun-Earth contrast ratio in the Johnson N band \citep{Ducati2001}. Surprisingly enough, it may thus 
be easier for the IR interferometer concept to detect a habitable Earth around an M-Dwarf than around an F star. 

Free flying interferometric systems like Darwin and TPF-I can be adapted to each individual target system. The baseline of 
the interferometers have to increase to resolve M star planetary systems but this has already been taken into account for the current 
 designs. Contrary to the interferometers, the current design of different coronographic concepts will 
have a minimum inner working angle due to the mask, favoring a wider  separation between planet and host star like e.g. 
for F stars. The inner working angle translates into a minimum separation needed between a planet and its host star in order 
for the former to be detectable. This selects against M dwarf stars in such a way that if these targets turn out to be very important a 
(nulling) interferometer is the only viable option.

\subsubsection{Early type stars}

 In any kind of unbiased survey, it is necessary to be extremely careful in the selection. In the context of 
 missions searching for worlds like our own, this has led to a number of different criteria  
 \citep{Turnbull2003, Brown2006,Mello2006}, being applied. Usually, when the goal is habitability, the studies focus on 
 solar type stars, including or excluding  M-dwarfs depending on the arguments for and against (see previous section), but 
 essentially always excluding early type stars. The argument for this is of course the short main-sequence life time, being 
 of the order of 500 million years for an early A-type star. Here we put forward a somewhat different argument. The number of 
 early type stars within the volume searchable with contemplated instruments is relatively low. Statistically, they will 
 therefore not allow anything to be said of eventual {\it non-detections}. On the other hand, the time span for the origin 
 of life on the Earth was maybe as short as 10 million years. This means that it would be both interesting and -- because of 
 the brightness \footnote{Because of the brightness of stars earlier than type F, the HZ is more distant from the star, and the 
 suppression of the stellar light becomes easier \citep{Kaltenegger2006}} and low numbers available within the search 
 volume -- not very time consuming to include them into the catalogue. We therefore suggest that stars earlier than class 
 F can be introduced into the search catalogue. We have included these stars in Table 3.

\subsection{Hosts to EGP}
Within 30 pc, 55 stars are known to host EGP with orbits that bring them within the field of view  of 
the Darwin interferometer and are part of the DASSC. Of these, 37 are part of the single target star sample, 
36 stars are part of the Emma target star sample, 25 of the $\pm$ 45\adeg cone design, and 10 of the 75mas 
inner working angle design. Some of these systems can be used to calibrate instruments - especially for 
EGP whose effective temperature can be determined with e.g. Spitzer already now. 
In addition characterizing these EGP systems should bring interesting results and allow the understanding of planetary 
systems with giant planets, and thus give clues to formation and composition of EGP. Several groups are 
working on the dynamical stability of terrestrial planets in the HZ in those systems (see e.g. \cite{Lunine2001}, 
\cite{Raymond2006}, \cite{Lohinger2006}), making some of the systems targets for habitable terrestrial planet search. 
Dynamical stability, formation scenarios and atmosphere calculations can be verified by observation of such planetary systems 
with the Darwin mission.

 \subsubsection{Number of stars required}
It is essential to have a certain number of targets stars to derive conclusions for detections or non-detections. 
This is a key problem when designing an instrument that can detect and study terrestrial exoplanets. 
The ESA Terrestrial Exoplanet Science Advisory Team (TE-SAT) have come to the conclusion that 
(Fridlund et al., 2008, in preparation) -- while just one detection of an 
Earth analogue, with or without the detection of biomarkers would constitute a remarkable result -- the desired result is 
15 detected systems with terrestrial analogues. This number would allow conclusions to be drawn with respect to the boundary 
conditions under which rocky planets are formed in the inner parts of planetary systems, the boundary conditions for formation 
and evolution, the actual evolutionary pathways -- at least in a rudimentary way -- and finally increase the chance of being 
able to discuss habitability. If we assume a 10\% planet fraction similar to the observed EGP fraction, a minimum of 150 stars need
to be observed to detect 15 terrestrial planets. Even though the frequency of extra-solar terrestrial planets can only be guessed at 
the moment, we believe that a sample has to consist of at least $>$ 150 stars in order to be able to derive conclusions -- 
particularly as what concerns non-detections.

\section{Conclusions}

In preparation for space missions searching for terrestrial planets among nearby stars we have established a target star 
catalogue of such objects. The list -- designated the Darwin  All Sky Star Catalogue or DASSC -- was created starting from the 
Hipparcos catalogue, and then taking into account other relevant catalogues. It currently contains a sample of 2303 identified 
objects of which 284 are F, 464 G, 883 K , 615 M type stars and 57 stars without B-V index.
Of these objects 949 objects are flagged in the DASSC as multiple using Hipparcos, CCDM and SB9.
185 of the objects are giant stars and 99 are flagged as dwarf stars, and the remaining 1962 stars have 733 multiple systems leading to a final DASSC sample of 1229 single main sequence stars,
(107 F stars, 235 G stars, 536 K stars and 351 M stars), where the size of the bolometric Habitable Zones have been identified. We have discussed elements of the target selection and the completeness of the DASSC, as well as the influence of factors like multiplicity and the presence of 
Extra solar Giant Planet host stars.
We have also shown how from the DASSC one can derive configuration dependent sub-catalogues. We have done this for three technical designs, the initial baseline design, 
the advanced Emma design as well as a catalogue using an inner working angle cut off. 
The DASSC will be the starting point for a multi-year effort studying the stars from all possible aspects with ground- and space-based assets. 

At the time of launch of a mission like Darwin or either of the TPF interferometer or coronographic versions, a prioritized list taking into consideration constraints imposed by the actual 
flight configuration, will be prepared based on the 
DASSC and the knowledge gained from a ground-based effort. 
The full catalogue can be obtained from one of the authors (LK) by sending an e-mail to lkaltene@cfa.harvard.edu.

\begin{deluxetable}{lllllllllll@{}llllllll} 
\tabletypesize{\scriptsize}
\rotate
\tablecaption{Sample and structure of the DASSC Part 1. The table in its entirety is available by sending an e-mail to lkaltene@cfa.harvard.edu}
\tablewidth{0pt}
\tablehead{
\colhead{DASSC}& \colhead{HIP} & \colhead{HD} & \colhead{d} & \colhead{$\Delta(d)$} &
 \colhead{B-V} & \colhead{V} & \colhead{Ecl long} & \colhead{Ecl lat} &
\colhead{$ccdm/SB9$} & \colhead{Flag} & \colhead{J} & \colhead{$\Delta_J$} & \colhead{H} & \colhead{$\Delta_H$} & \colhead{K} & 
\colhead{$\Delta_K$} & \colhead{Qflg} & \colhead{Sp}\\
&&&[pc]&[pc]&[mag]&[mag&[deg]&[deg]&   &&[mag]&[mag]&[mag]&[mag]&[mag]&[mag]&

%HOW TO ADD LINE WITH UNITS BELOW??????

%\colhead{} & \colhead{} & \colhead{} & \colhead{[pc]} & \colhead{[pc]} & \colhead{[mag]} & \colhead{[mag]} & \colhead{[deg]}  & \colhead{[deg]}  & \colhead{[deg]} & \colhead{[deg]} & \colhead{} & \colhead{HIP} & \colhead{[mag]} & \colhead{[mag]} & 
%\colhead{[mag]} & \colhead{[mag]} & \colhead{[mag]} & \colhead{[mag]} & \colhead{} & \colhead{}

}
\startdata
1	&	70890	&		&	1.3	&	0.00	&	1.807	&	11.01	&	239.11465	&	-44.76331	&	14396-6050	&	 	&	5.357	&	0.023	&	4.835	&	0.057	&	4.384	&	0.033	&	AAE	&	M5Ve        	\\																																																																																					
2	&	71683	&	128620	&	1.3	&	0.00	&	0.71	&	-0.01	&	239.47932	&	-42.59432	&	14396-6050	&	C	&	-1.454	&	0.133	&	-1.886	&	0.22	&	-2.008	&	0.26	&	BDD	&	G2V         	\\																																																																																					
3	&	71681	&	128621	&	1.3	&	0.00	&	0.9	&	1.35	&	239.47756	&	-42.59841	&	14396-6050	&	C	&	-1.454	&	0.133	&	-1.886	&	0.22	&	-2.008	&	0.26	&	BDD	&	K1V         	\\																																																																																					
4	&	54035	&	95735	&	2.5	&	0.01	&	1.502	&	7.49	&	152.10713	&	27.39394	&	11033+3558	&	 	&	4.203	&	0.242	&	3.64	&	0.202	&	3.254	&	0.306	&	DCD	&	M2V         	\\																																																																																					
5	&	16537	&	22049	&	3.2	&	0.01	&	0.881	&	3.72	&	48.16776	&	-27.71577	&	03329-0927	&	 	&	2.228	&	0.298	&	1.88	&	0.276	&	1.776	&	0.286	&	DDD	&	K2V         	\\																																																																																					
6	&	114046	&	217987	&	3.3	&	0.01	&	1.483	&	7.35	&	332.68430	&	-27.51180	&	          	&	 	&	4.338	&	0.258	&	3.608	&	0.23	&	3.465	&	0.2	&	DDC	&	M2/M3V      	\\																																																																																					
7	&	57548	&		&	3.3	&	0.02	&	1.746	&	11.12	&	176.86783	&	-0.48035	&	          	&	 	&	6.505	&	0.023	&	5.945	&	0.024	&	5.654	&	0.024	&	AAA	&	M4.5V       	\\																																																																																					
8	&	104214	&	201091	&	3.5	&	0.02	&	1.069	&	5.2	&	336.95686	&	51.89908	&	21069+3844	&	 	&	3.114	&	0.268	&	2.54	&	0.198	&	2.248	&	0.318	&	DCD	&	K5V         	\\																																																																																					
9	&	37279	&	61421	&	3.5	&	0.01	&	0.432	&	0.4	&	115.78553	&	-16.01962	&	07393+0514	&	O	&	-0.498	&	0.151	&	-0.666	&	0.27	&	-0.658	&	0.322	&	BDD	&	F5IV-V      	\\																																																																																					
10	&	104217	&	201092	&	3.5	&	0.01	&	1.309	&	6.05	&	336.95741	&	51.89057	&	21069+3844	&	 	&	3.546	&	0.278	&	2.895	&	0.218	&	2.544	&	0.328	&	DDD	&	K7V         	\\																																																																																					
11	&	91772	&	173740	&	3.5	&	0.06	&	1.561	&	9.7	&	309.79113	&	81.56859	&	18428+5937	&	C	&	5.721	&	0.02	&	5.197	&	0.024	&	5	&	0.023	&	AAA	&	K5          	\\																																																																																					
12	&	1475	&	1326	&	3.6	&	0.01	&	1.56	&	8.09	&	24.67504	&	37.92904	&	00184+4401	&	 	&	5.252	&	0.264	&	4.476	&	0.2	&	4.018	&	0.02	&	DCE	&	M1V         	\\																																																																																					
13	&	108870	&	209100	&	3.6	&	0.01	&	1.056	&	4.69	&	309.62702	&	-41.40882	&	          	&	 	&	2.894	&	0.292	&	2.349	&	0.214	&	2.237	&	0.24	&	DCD	&	K5V         	\\																																																																																					
14	&	8102	&	10700	&	3.6	&	0.01	&	0.727	&	3.49	&	17.81860	&	-24.81586	&	01441-1557	&	 	&	2.149	&	0.31	&	1.8	&	0.234	&	1.794	&	0.274	&	DDD	&	G8V         	\\																																																																																					
15	&	36208	&		&	3.8	&	0.02	&	1.573	&	9.84	&	112.74273	&	-16.50485	&	          	&	 	&	5.714	&	0.032	&	5.219	&	0.063	&	4.857	&	0.023	&	AAA	&	M5          	\\																																																																																					
16	&	24186	&	33793	&	3.9	&	0.01	&	1.543	&	8.86	&	67.25093	&	-67.50597	&	          	&	 	&	5.821	&	0.026	&	5.316	&	0.027	&	5.049	&	0.021	&	AAA	&	M0V         	\\																																																																																					
17	&	105090	&	202560	&	3.9	&	0.02	&	1.397	&	6.69	&	309.53560	&	-21.95219	&	          	&	 	&	4.046	&	0.266	&	3.256	&	0.216	&	3.1	&	0.23	&	DCD	&	M1/M2V      	\\																																																																																					
18	&	110893	&	239960	&	4.0	&	0.05	&	1.613	&	9.59	&	16.38193	&	59.15366	&	22281+5741	&	C	&	5.575	&	0.027	&	5.038	&	0.034	&	4.777	&	0.029	&	AAA	&	M2V         	\\																																																																																					
19	&	30920	&		&	4.1	&	0.04	&	1.69	&	11.12	&	98.17345	&	-26.04504	&	06294-0249	&	G	&	6.376	&	0.023	&	5.754	&	0.034	&	5.486	&	0.016	&	AAA	&	M4.5Ve      	\\																																																																																					
20	&	80824	&		&	4.3	&	0.03	&	1.604	&	10.1	&	247.85797	&	9.06959		&	          	&	 	&	5.95	&	0.024	&	5.373	&	0.04	&	5.075	&	0.024	&	AAA	&	M4          	\\																																																																													
\enddata
%% Text for table notes should follow after the \enddata but before
%% the \end{deluxetable}. Make sure there is at least one \tablenotemark
%% in the table for each \tablenotetext.
%
\end{deluxetable}

\begin{deluxetable}{lllllllllllll} 
\tabletypesize{\scriptsize}
\rotate
\tablecaption{Sample and structure of the DASSC part 2. The table in its entirety is available by sending an e-mail to lkaltene@cfa.harvard.edu}
\tablewidth{0pt}
\tablehead{
\colhead{DASSC} & \colhead{HIP} & \colhead{HD} & \colhead{L}& \colhead{Radius} &
 \colhead{$T_{eff}$} &  \colhead{$HZ_{R}$} & \colhead{$HZ_{R}$} &
\colhead{$HZ_{I}$} & \colhead{$HZ_{O}$} & \colhead{$V-K$} &\colhead{$K_{J}$}&\colhead{$BC$}\\
&&&[$L_{sun}$]&[$R_{sun}$]&[K] &[AU] &[mas]& [AU] &[AU] &[AU] &[mag] & [mag]
}
\startdata
1	&	70890	&	 	&	0.10	&	0.03	&	2687	&	0.45	&	3.49E+02	&	0.30	&	0.60	&	6.582	&	4.428	&	-8.18	\\																																																																																																									
2	&	71683	&	128620	&	1.84	&	1.54	&	5098	&	1.70	&	1.26E+03	&	1.14	&	2.26	&	1.954	&	-1.964	&	-0.265	\\																																																																																																									
3	&	71681	&	128621	&	1.17	&	1.34	&	3994	&	1.57	&	1.16E+03	&	1.06	&	2.08	&	3.314	&	-1.964	&	-1.132	\\																																																																																																									
4	&	54035	&	95735	&	0.04	&	0.19	&	3512	&	0.30	&	1.16E+02	&	0.20	&	0.39	&	4.192	&	3.298	&	-2.261	\\																																																																																																									
5	&	16537	&	22049	&	0.33	&	0.65	&	5159	&	0.83	&	2.59E+02	&	0.56	&	1.11	&	1.900	&	1.82	&	-0.241	\\																																																																																																									
6	&	114046	&	217987	&	0.05	&	0.24	&	3687	&	0.32	&	9.72E+01	&	0.22	&	0.42	&	3.841	&	3.509	&	-1.742	\\																																																																																																									
7	&	57548	&	 	&	0.03	&	0.06	&	3026	&	0.23	&	7.02E+01	&	0.16	&	0.31	&	5.422	&	5.698	&	-4.804	\\																																																																																																									
8	&	104214	&	201091	&	0.17	&	0.52	&	4258	&	0.59	&	1.70E+02	&	0.40	&	0.79	&	2.908	&	2.292	&	-0.807	\\																																																																																																									
9	&	37279	&	61421	&	6.62	&	2.03	&	6511	&	2.83	&	8.09E+02	&	1.87	&	3.79	&	1.014	&	-0.614	&	0.007	\\																																																																																																									
10	&	104217	&	201092	&	0.12	&	0.42	&	3895	&	0.50	&	1.44E+02	&	0.34	&	0.67	&	3.462	&	2.588	&	-1.296	\\																																																																																																									
\enddata
%% Text for table notes should follow after the \enddata but before
%% the \end{deluxetable}. Make sure there is at least one \tablenotemark
%% in the table for each \tablenotetext.
%\tablenotetext{a}{ Luminosity (L), effective temperature ($T_{eff}$), Radius of the Habitable Zone (HZ) ($HZ_{R}$), Inner Edge of the HZ	($HZ_{I}$), Outer Edge of the HZ ($HZ_{O}$), Johnson K Magnitude ($K_{J}$) }
\end{deluxetable}
\begin{deluxetable}{lllllllllllllllllll} 
\tabletypesize{\scriptsize}
\rotate
\tablecaption{Characteristics of Giants and Dwarf stars in the DASSC. The table in its entirety is available by sending an e-mail to lkaltene@cfa.harvard.edu}
\tablewidth{0pt}
\tablehead{
\colhead{HIP} & \colhead{HD} & \colhead{d} & \colhead{$\Delta d$$\tablenotemark{a}$} &
 \colhead{B-V} & \colhead{V} & \colhead{Ecl long} & \colhead{Ecl lat} &
  \colhead{J} & \colhead{$\Delta_J$} & \colhead{H} & \colhead{$\Delta_H$} & \colhead{K} & \colhead{$\Delta_K$} & \colhead{Sp}\\
&&[pc]&[pc]&[mag]&[mag&[deg]&[deg]&[mag]&[mag]&[mag]&[mag]&[mag]&[mag]&
}

\startdata
62956	&	112185	&	24.81	&	0.38	&	-0.022	&	1.76	&	158.93	&	54.32          	&		1.713	&	0.246	&	1.698	&	0.186	&	1.625	&	0.266	&	A0p         	\\																																																																																																																																																									
84143	&	155203	&	21.95	&	0.37	&	0.441	&	3.32	&	260.74	&	-20.18	& 2.398	&	0.272	&	2.295	&	0.236	&	2.255	&	0.252	&	F3p         	\\																																																																																																																																																									
32362	&	48737	&	17.54	&	0.25	&	0.443	&	3.35	&	101.21	&	-10.10	&  1.996	&	0.23	&	1.811	&	0.168	&	1.688	&	0.22	&	F5IV        	\\																																																																																																																																																							
69701	&	124850	&	21.39	&	0.39	&	0.511	&	4.07	&	213.80	&	7.20	&	3.14	&	0.266	&	2.909	&	0.236	&	2.801	&	0.266	&	F7V         	\\																																																																																																																																																						
36795	&	60532	&	25.70	&	0.43	&	0.521	&	4.44	&	120.47	&	-43.28	& 3.703	&	0.27	&	3.318	&	0.24	&	3.355	&	0.286	&	F6V         	\\																																																																																																																																																									
78527	&	144284	&	20.92	&	0.23	&	0.528	&	4.01	&	196.67	&	74.44	& 2.589	&	0.278	&	2.394	&	0.21	&	2.156	&	0.26	&	F8IV-V      	\\																																																																																																																																																							
74605	&	136064	&	25.31	&	0.30	&	0.55	&	5.15	&	158.20	&	74.10	& 4.217	&	0.246	&	3.983	&	0.248	&	3.872	&	0.356	&	F9IV  \\																																																																																																																																																									
2021	&	2151	&	7.47	&	0.03	&	0.618	&	2.82	&	300.97	&	-64.78	&	1.676	&	0.296	&	1.434	&	0.22	&	1.372	&	0.248	&	G2IV        	\\																																																																																																																																																					
22336	&	30562	&	26.50	&	0.61	&	0.631	&	5.77	&	69.82	&	-27.87	&	4.984	&	0.262	&	4.574	&	0.266	&	4.31	&	0.049	&	F8V         	\\																																																																																																																																																									
7276	&	9562	&	29.66	&	0.62	&	0.639	&	5.75	&	18.99	&	-15.62	& 4.627	&	0.029	&	4.391	&	0.015	&	4.258	&	0.018	&	G2IV        	\\																																																																																																																																																						
\enddata
%% Text for table notes should follow after the \enddata but before
%% the \end{deluxetable}. Make sure there is at least one \tablenotemark
%% in the table for each \tablenotetext.
\tablenotetext{a}{Estimated error in distance}
\end{deluxetable}

%UPDATE

\begin{deluxetable}{llllllllllrllllllll} 
\tabletypesize{\scriptsize}
\rotate
\tablecaption{DASSC EGP Host stars and system parameters. The table in its entirety is available by sending an e-mail to lkaltene@cfa.harvard.edu}
\tablewidth{0pt}
\tablehead{
\colhead{Star}	 & \colhead{Planet Name} & \colhead{$M_{pl}$} & \colhead{Period} & \colhead{Sem Maj} & \colhead{Ecc} & \colhead{Incl} & \colhead{Ang.Dist} & \colhead{d} & \colhead{Spec.Type} & \colhead{$[Fe/H]$}& \colhead{RAJ2000} & \colhead{DEJ2000} & \colhead{V} & \colhead{$R_{Star}$}\\
& &[$M_{Jup}$] & [days] & [AU] &  & [deg] & [as] & [pc] & &  &  & & [mag] & [$R_{Sol}$]
}
\startdata
HD 101930	&	HD 101930 b	&	0.3	&	70.46	&	0.302	&	0.11	&		&	0.01	&	30.49	&	K1 V	&		0.17	&	11 43 30	&	-58 00 24	&	8.21	&	\\																																						
HD 102195	&	HD 102195 b	&	0.45	&	4.11377	&	0.049	&	0	&		&	0.00169	&	28.98	&	K0V	&		0.05	&	11 45 42	&	+02 49 17	&	8.05	&	0.835\\																						
HD 10647	&	HD 10647 b	&	0.91	&	1040	&	2.1	&	0.18	&		&	0.121	&	17.3	&	F8V	&		-0.03	&	01 42 29	&	-53 44 27	&	5.52	&	1.1\\																																						
HD 111232	&	HD 111232 b	&	6.8	&	1143	&	1.97	&	0.2	&		&	0.068	&	29	&	G8V	&		-0.36	&	12 48 51	&	-68 25 30	&	7.61	&	\\																							HIP 113020	&	GJ 876 b	&	1.94	&	60.9	&	0.21	&	0.02	&	84	&	44.0	&	4.7	&	M4V	&	-0.12	&	22 53 13	&	-14 15 13	&	10.17	&	0.4	\\
HIP 113020	&	GJ 876 c	&	0.56	&	30.1	&	0.13	&	0.27	&	84	&	27.5	&	4.7	&	M4V	&	-0.12	&	22 53 13	&	-14 15 13	&	10.17	&	0.4	\\
HIP 113020	&	GJ 876 d	&	0.02	&	1.9	&	0.02	&	0.00	&		&	4.4	&	4.7	&	M4V	&	-0.12	&	22 53 13	&	-14 15 13	&	10.17	&	0.4	\\														
HD 114386	&	HD 114386 b	&	0.99	&	872	&	1.62	&	0.28	&		&	0.058	&	28	&	K3 V	&		-0.03	&	13 10 39	&	-35 03 17	&	8.73	&	0.76 \\																																						
HD 114783	&	HD 114783 b	&	0.99	&	501	&	1.2	&	0.1	&		&	0.055	&	22	&	K0	&		0.33	&	13 12 43	&	-02 15 54	&	7.57	&	0.78	\\																																						
HD 120136	&	Tau Boo b	&	3.9	&	3.3135	&	0.046	&		&		&	0.00306	&	15	&	F7 V	&		0.28	&	13 47 17	&	+17 27 22	&	4.5	&	0.84\\																					
HD 128311	&	HD 128311 b	&	2.18	&	448.6	&	1.099	&	0.25	&		&	0.06620	&	16.6	&	K0	&		0.08	&	14 36 00	&	+09 44 47	&	7.51	& 0.73\\																			
HD 128311	&	HD 128311 c	&	3.21	&	919	&	1.76	&	0.17	&		&	0.10602	&	16.6	&	K0	&		0.08	&	14 36 00	&	+09 44 47	&	7.51	&	0.73	\\																																					
HD 130322	&	HD 130322 b	&	1.08	&	10.724	&	0.088	&	0.048	&		&	0.003	&	30	&	K0 V	&		-0.02	&	14 47 32	&	-00 16 53	&	8.05	&	0.83	\\																																						
HD 134987	&	HD 134987 b	&	1.58	&	260	&	0.78	&	0.24	&		&	0.031	&	25	&	G5 V	&		0.23	&	15 13 28	&	-25 18 33	&	6.45	&	1.2	\\																																						
HD 142		&	HD 142 b	&	1	&	337.112	&	0.98	&	0.38	&		&	0.048	&	20.6	&	G1 IV	&		0.04	&	00 06 19	&	-49 04 30	&	5.7	&	0.86	\\																																						
\enddata
%% Text for table notes should follow after the \enddata but before
%% the \end{deluxetable}. Make sure there is at least one \tablenotemark
%% in the table for each \tablenotetext.

\end{deluxetable}

\acknowledgments
\begin{acknowledgements}
Special thanks to G Torres and E. Mamajek for many stimulating discussion and very constructive suggestions. 
Part of this work was supported by NASA grant NAG5-13045. This research has made use of the SIMBAD database,
operated at CDS, Strasbourg, France, NASA's Astrophysics Data System and data products from the Two Micron All Sky Survey, 
which is a joint project of the UMass and IPAC, funded by NASA and NSF.
\end{acknowledgements}

\bibliographystyle{Spr-mp-nameyear-cnd}
\bibliography{LKFridlund_oct21}
\end{document}